\documentclass[twocolumn,preprint]{aastex631}

\usepackage{placeins}
\usepackage{subfigure}
\usepackage{amsmath}

\usepackage{longtable}
\usepackage{booktabs}
\usepackage{natbib}
\usepackage{tabularx}
\usepackage{graphicx}
\usepackage{tikz}
\usetikzlibrary{arrows.meta, positioning}
\usepackage{pstricks}
\usepackage{hyperref}
\usepackage{pst-node}
\usepackage{array} 
\received{May 13, 2025}
\revised{June 17, 2025}
\accepted{June 18, 2025}
\journalinfo{Accepted to ApJL on June 18, 2025}
\definecolor{mat}{HTML}{000064}

\begin{document}
\title{VLT/ERIS observations of the V960 Mon system: a dust-embedded substellar object formed by gravitational instability?}
\author[0009-0009-8115-8910]{Anuroop Dasgupta}
\affiliation{European Southern Observatory, Alonso de Córdova 3107, Casilla
19, Santiago, Chile}
\affiliation{Instituto de Estudios Astrof\'isicos, Facultad de Ingenier\'ia y Ciencias, Universidad Diego Portales, Av. Ej\'ercito Libertador 441, Santiago, Chile}
\affiliation{Millennium Nucleus on Young Exoplanets and their Moons (YEMS), Chile}
\email{anuroop.dasgupta@mail.udp.cl}

\author[0000-0002-5903-8316]{Alice Zurlo}
\affiliation{Instituto de Estudios Astrof\'isicos, Facultad de Ingenier\'ia y Ciencias, Universidad Diego Portales, Av. Ej\'ercito Libertador 441, Santiago, Chile}
\affiliation{Millennium Nucleus on Young Exoplanets and their Moons (YEMS), Chile}

\author[0000-0002-3354-6654]{Philipp Weber}
\affiliation{Departamento de Física, Universidad de Santiago de Chile, Av. Victor Jara 3659, Santiago, Chile}
\affiliation{Millennium Nucleus on Young Exoplanets and their Moons (YEMS), Chile}
\affiliation{Center for Interdisciplinary Research in Astrophysics and Space Exploration (CIRAS), Universidad de Santiago de Chile, Chile}

\author[0000-0002-6395-8916]{Francesco Maio}
\affiliation{INAF–Osservatorio Astrofisico di Arcetri, Largo E. Fermi 5, 50125 Firenze, Italy}
\affiliation{Universit\`a di Firenze, Dipartimento di Fisica e Astronomia, Via Giovanni Sansone 1, 50019 Sesto Fiorentino FI, Italy}

\author[0000-0002-2828-1153]{Lucas A. Cieza}
\affiliation{Instituto de Estudios Astrof\'isicos, Facultad de Ingenier\'ia y Ciencias, Universidad Diego Portales, Av. Ej\'ercito Libertador 441, Santiago, Chile}
\affiliation{Millennium Nucleus on Young Exoplanets and their Moons (YEMS), Chile}

\author[0000-0001-6156-0034]{Davide Fedele}
\affiliation{INAF–Osservatorio Astrofisico di Arcetri, Largo E. Fermi 5, 50125 Firenze, Italy}

\author[0000-0002-4266-0643]{Antonio Garufi}
\affiliation{INAF–Istituto di Radioastronomia, Via Gobetti 101, I-40129 Bologna, Italy}
\affiliation{Max-Planck-Institut für Astronomie, Königstuhl 17, 69117 Heidelberg, Germany}

\author[0000-0002-1575-680X]{James Miley}
\affiliation{Departamento de Física, Universidad de Santiago de Chile, Av. Victor Jara 3659, Santiago, Chile}
\affiliation{Millennium Nucleus on Young Exoplanets and their Moons (YEMS), Chile}
\affiliation{Center for Interdisciplinary Research in Astrophysics and Space Exploration (CIRAS), Universidad de Santiago de Chile, Chile}

\author[0000-0003-3991-6107]{Prashant Pathak}
\affiliation{Department of SPASE, Indian Institute of Technology, Kanpur, India}

\author[0000-0002-5772-8815]{Sebastián Pérez}
\affiliation{Departamento de Física, Universidad de Santiago de Chile, Av. Victor Jara 3659, Santiago, Chile}
\affiliation{Millennium Nucleus on Young Exoplanets and their Moons (YEMS), Chile}
\affiliation{Center for Interdisciplinary Research in Astrophysics and Space Exploration (CIRAS), Universidad de Santiago de Chile, Chile}

\author[0000-0002-4650-594X]{Veronica Roccatagliata}
\affiliation{Alma Mater Studiorum, Università di Bologna, Dipartimento di Fisica e Astronomia (DIFA), Via Gobetti 93/2, 40129 Bologna, Italy}
\affiliation{INAF–Osservatorio Astrofisico di Arcetri, Largo E. Fermi 5, 50125 Firenze, Italy}

\begin{abstract}
V960~Mon is an FU Orionis object that shows strong evidence of a gravitationally unstable spiral arm that is fragmenting into several dust clumps. We report the discovery of a new substellar companion candidate around this young star, identified in high-contrast $L'$-band imaging with VLT/ERIS. The object is detected at a projected separation of $0.898 \pm 0.01\arcsec$ with a contrast of $(8.39 \pm 0.07) \times 10^{-3}$. The candidate lies close to the clumps previously detected in the sub-mm  (at 1.3 mm) and is co-located with extended polarized IR signal from scattered stellar irradiation, suggesting it is deeply embedded. The object is undetected in the SPHERE $H$-band total intensity, placing an upper mass limit of $\sim38~M_\mathrm{Jup}$ from the contrast curve. Using evolutionary models at an assumed age of 1~Myr, we estimate a mass of $\sim660~M_\mathrm{Jup}$ from the L' brightness; however, this value likely includes a significant contribution from a disk around the companion. The discrepancy between near- and mid-infrared results again suggests the source is deeply embedded in dust. This candidate may represent an actively accreting, disk-bearing substellar object in a young, gravitationally unstable environment.

\end{abstract}

\keywords{
Direct imaging,
Exoplanet detection methods,
Exoplanet astronomy,
Exoplanets,
Planet formation,
Protoplanetary disks,
Protostars,
Star formation,
Planetary-disk interactions,
Adaptive optics,
Astronomy image processing,
High contrast techniques
}

\section{Introduction} \label{sec:intro}

FU Orionis objects (FUors) are young stars that exhibit dramatic luminosity outbursts. The sharp increase in brightness is caused by episodic accretion events that significantly alter their circumstellar environment \citep[see reviews by][]{Audard2014, Fischer2023}. The mechanism that invokes these accretion bursts is still actively discussed. Many works connect the transient accretion event with gravitational instability \citep[GI,][]{Toomre1964} occurring in the outer protoplanetary disk \citep[e.g.][]{Armitage2001, Zhu2009, Vorobyov2020}, making FUors preferential targets to search for features of GI. Tracing circumstellar environments for such signatures is of utmost interest, as planet formation by GI \citep{Boss1997} is discussed as the preferred mechanism for explaining giant planets at large separation from their host stars in exoplanet statistics \citep{2016ARA&A..54..271K,Nielsen2019,2021A&A...651A..72V}. However, observational evidence of gravitationally unstable environments is so far limited to very few cases, including Elias~2-27 \citep{Paneque2021}, V960~Mon \citep{Weber2023}, and AB~Aurigae \citep{Speedie2024}. Out of those, only V960~Mon represents an FUor, undergoing an outburst since 2014 \citep{Maehara2014}. 
Thus, besides the putative opportunity to study ongoing GI, it presents an excellent target for studying the environment of an outburst at an early stage.

Gravitational instability is not only invoked to explain transient accretion episodes in FUor systems but also represents a fundamental pathway for the formation of bound companions within massive, young circumstellar disks \citep{Boss1997, 2010ApJ...710.1375K}. In this scenario, the disk becomes locally unstable and fragments into self-gravitating clumps, which may collapse to form substellar or stellar-mass companions \citep[e.g.,][]{Boss1997,2010ApJ...710.1375K}. While such fragmentation is well-supported by numerous numerical simulations \citep[e.g.,][]{2010ApJ...719.1896V,2024arXiv241012042X,2024A&A...682L...6F}, direct observational evidence of substellar companion formation through gravitational instability remains elusive.
 The only robust case of disk fragmentation confirmed to date involves a stellar companion in the protostellar system L1448 IRS3B, where ALMA observations revealed spiral arms and a compact source consistent with a forming star embedded within a gravitationally unstable disk \citep{2016Natur.538..483T}. However, no substellar companion forming through this mechanism has yet been conclusively observed. Identifying such a case would provide the first direct evidence of planet or brown dwarf formation via GI, offering a crucial test of theoretical models and advancing our understanding of the early stages of companion formation in massive disks.

The presence of spiral structures and clump-like features in ALMA observations of the V960~Mon system \citep{Weber2023} makes it a particularly compelling candidate for investigating the link between accretion variability, disk instability, and potential companion formation \citep{ Vorobyov2020}. Moreover, companions themselves may play a critical role in triggering or sustaining accretion events in FUor systems \citep{2004ApJ...608L..65R,2004MNRAS.353..841L}, adding to the motivation to search for faint objects within their disks.

Recent advances in high-contrast imaging (HCI) have opened new avenues for detecting faint substellar companions and circumplanetary structures in young systems \citep{2016PASP..128j2001B,2018A&A...617A..44K,2019A&A...632A..25M,2021SPIE11823E..04C,2024arXiv240405797Z}. Observations in the thermal infrared, particularly in the $L'$-band, are well suited for tracing warm, embedded objects that may be obscured at shorter wavelengths. The Enhanced Resolution Imager and Spectrograph (ERIS) at the Very Large Telescope (VLT) provides diffraction-limited imaging capabilities in this regime and is optimized for detecting faint sources at small angular separations \citep{Davies2023}. \citet{Maio2025} demonstrated the capability of ERIS with the vAPP coronagraph to resolve infrared disk structures in several protoplanetary systems at 4\,$\mu$m. Their observations revealed gaps, spiral arms, and scattered light features across multiple disks, highlighting ERIS’s effectiveness for probing disk morphology and setting constraints on planet formation across a range of environments.

Finally, recent theoretical models suggest that circumplanetary disks (CPDs) can significantly contribute to the infrared and line emission of young forming companions \citep{2015ApJ...799...16Z,2016ApJ...832..193Z,Taylor2025}. Radiation-hydrodynamic simulations by \citet{2018ApJ...866...84A} show that accretion shocks at the surface of protoplanets and within their CPDs can produce strong hydrogen emission lines such as H$\alpha$, Pa$\beta$, and Br$\gamma$. These accretion-driven processes not only trace ongoing mass infall but can also substantially add to the thermal emission of the system, increasing the total flux, particularly at near- and mid-infrared wavelengths. Consequently, the observed luminosity of young substellar objects may be significantly elevated by circumplanetary material, leading to overestimated mass values when interpreted solely through evolutionary models. 

In this study, we present high-contrast, high-angular resolution $L'$-band observations of the FUor system V960~Mon using VLT/ERIS, complemented by archival SPHERE and ALMA data. This paper is structured as follows: in Section~\ref{obs and data reduction}, we describe the observations and data reduction procedures; in Section~\ref{sec:results}, we present the imaging analysis, photometry, and mass estimates; in Section~\ref{sec:discussion}, we interpret the findings in the context of companion formation and circumplanetary emission; and in Section~\ref{sec:conclusions}, we summarize our conclusions and discuss future prospects.

\section{Observations and Data Reduction}\label{obs and data reduction}
\subsection{Observations}\label{sec:obs}

We observed V960~Mon using the Enhanced Resolution Imager and Spectrograph (ERIS; \citealt{Davies2023}) mounted on UT4 of the Very Large Telescope (VLT), as part of program 112.25PY.001 (PI: P. Weber). Observations were carried out on January 1st, 2024, using the Near Infrared Camera System (NIX) in imaging mode with the $L'$-band filter ($\lambda_{\rm obs} = 3.8\,\mu$m). The NIX detector has known defects in its lower two quadrants, including bad pixels and irregular pixel response, which can compromise image quality. To avoid these issues, we placed the target in the upper two quadrants of the detector where the pixel behavior is more stable \citep[see][]{Maio2025}.

We employed a two-position dithering pattern to minimize static detector noise and facilitate background subtraction. The star was jittered between the upper quadrants during the sequence. Observations were taken in field-stabilized mode (without using a coronagraph), under very good atmospheric conditions. The median seeing during the observations was at $\sim$0.47\arcsec, with an atmospheric coherence time $\tau_0 \approx 9.8$\,ms and wind speeds were measured at 4.8\,m/s and 5.1\,m/s at 10\,m and 30\,m heights, respectively. 

The observing sequence consisted of 16 data cubes, each comprising 500 frames taken with an integration time (DIT) of 0.1s, resulting in a total exposure time of approximately 13 minutes.


\subsection{Data Reduction} \label{sec:data red}

Our data reduction followed standard pre-processing procedures, including dark subtraction, bad pixel correction via sigma-clipping, PSF centering using 2D Gaussian fitting, flat field normalization, local background subtraction, and the removal of horizontal and vertical detector artifacts. Edge effects were mitigated by masking regions near the detector border, and the final science image was obtained by stacking the processed frames from all cubes.

We fol reduction workflow presented in \citet{Maio2025}, although we emphasize that their study utilized the vAPP coronagraph, unlike our observations, which were conducted in direct imaging mode. 

To improve computational efficiency, we implemented a binning strategy by averaging every 10 consecutive frames within each cube, reducing the data volume by a factor of ten.

The nominal field of view (FoV) of the ERIS/NIX detector is $26.4\arcsec \times 26.4\arcsec$ when using the 13~mas/pixel plate scale \citep{Davies2023}. However, due to our observational strategy—which involved jittering between the two upper quadrants of the detector to avoid the known concentration of bad pixels in the lower regions,the effective angular coverage in the final reduced image is smaller. Specifically, our stacked science image corresponds to an angular FoV of approximately $11.7\arcsec \times 11.7\arcsec$.


\section{Results}\label{sec:results}

\begin{figure*}
        \centering
    \includegraphics[]{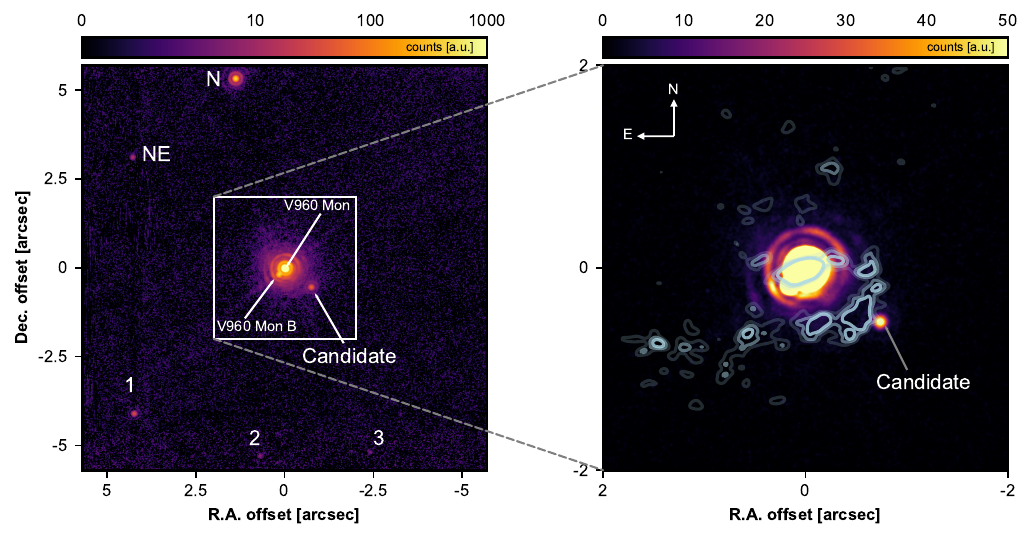}
    \caption{VLT/ERIS image of V960~Mon in $L'$-band. The left panel shows the FUor object embedded in its environment, marking the detection of V960~Mon~N and V960~Mon~NE. The right panel shows a zoom-in onto V960~Mon, overlayed with ALMA band~6 continuum contours at 3, 4, and 5 $\sigma_{\rm rms}$ \citep[cf. Fig.~2 in][]{Weber2023}.}
    \label{fig:ERIS}
\end{figure*}

\subsection{Candidate detection, astrometry and photometry}

The final reduced $L'$-band image from ERIS is shown in Figure~\ref{fig:ERIS}. Within the effective field of view, we detect several distinct point sources associated with the V960~Mon system. The central FUor, V960~Mon, is clearly detected. Its well-known northern companion, V960~Mon~N  (also known as UCAC4~430-024261), first identified by \citet{Kospal2015}, is also recovered. This source has since been confirmed at multiple wavelengths, including ALMA continuum emission \citep{2021ApJS..256...30K,Weber2025}.

A second previously known source, V960~Mon~NE, is detected to the northeast of the primary. This source was initially highlighted in \citet{Weber2025} using VLT/MUSE and SPHERE data. Although already marginally visible in earlier H$\alpha$ observations, it was not explicitly cataloged at that time. We recover it at the expected position in our ERIS $L'$-band image.

We also note a close-in companion (V960~Mon B) southeast of the primary star, located within the PSF core  (close to the first airy ring) of V960~Mon.  It was first identified in $J$- and $K$-bands by \citet{Caratti2015}, who reported strong Pa$\beta$ and Br$\gamma$ emission with a separation of $227$\,mas and PA of $131.4^{\circ}$. \citet{Weber2025} later confirmed this source in polarized and total intensity SPHERE images, reporting a slightly larger separation of $237 \pm 4$\,mas and a PA of $136.7^{\circ}$. Although not resolved here, the companion manifests as an elongation of the central PSF, as seen in Figure~\ref{fig:ERIS} (right).

In addition to the known sources, we detect three additional compact sources in our ERIS $L'$-band image, which we label as 1, 2, and 3. While these objects were not explicitly reported in earlier studies, they are faintly visible in the $JHK_s$ color-composite image presented by \citet{Kospal2015}, indicating that their presence had already been captured but not analyzed in detail. 

Importantly, the previously reported far-infrared source V960~Mon~SE \citep{Kospal2015,2021ApJS..256...30K} lies outside the effective field of view of our ERIS data and is therefore not recovered here.

Among the detected sources, the most notable is a previously unreported point-like object, which we refer to as the candidate, located southeast of the central star. This source, not cataloged in earlier studies, is detected with high significance in our ERIS $L'$-band image. Its clear detection motivates a detailed astrometric and photometric analysis to assess its potential as a new companion to V960~Mon.

For all identified sources—UCAC4~430-024261, NE, the unresolved southeast companion, the three additional detections (1, 2, and 3), and the new companion candidate, we derive relative astrometry and photometry. Fluxes, separations, position angles, and contrasts with respect to the central source are summarized in Table~\ref{tab:positions}.

Precise astrometric positions of the detected sources were measured relative to the central star. For the new companion candidate, we derive a relative position of $\Delta$RA = $-0\farcs745 \pm 0.002$ and $\Delta$Dec = $-0\farcs510 \pm 0.002$, corresponding to a projected separation of $0.898 \pm 0.009\arcsec$ and a position angle (PA) of $234.13^{\circ} \pm 0.01$. Astrometry was performed by fitting a two-dimensional Gaussian profile to the companion signal, after centering the frame using a Gaussian fit to the PSF core of the central star.

Photometry of the candidate was performed using aperture photometry with a radius of 15 pixels ($0.205\arcsec$), chosen to encompass the PSF while minimizing contamination from the surrounding stellar halo. The background level dominated by the bright diffraction features of V960~Mon was estimated from a surrounding annulus (20–30 pixels) and subtracted. This method effectively removes the local halo flux while avoiding bias from nearby clumps or detector artifacts. We measured a contrast of $(8.39 \pm 0.07)\times 10^{-3}$ for the companion relative to the central star.

Due to the lack of a dedicated PSF reference star in the ERIS observations, we cannot fully model or subtract the stellar halo through PSF fitting. To assess the potential systematic uncertainty from this limitation, we performed a series of aperture photometry tests by varying the aperture radius from 13–17 pixels and changing the annular background region. These tests yielded contrast values in the range $(8.2–8.6) \times 10^{-3}$, corresponding to a maximum deviation of $\sim2.5\%$ from our nominal contrast value. We therefore estimate the total uncertainty in the companion’s photometry including photon noise, background subtraction error, and systematic halo contribution to be $\sim0.07 \times 10^{-3}$ (i.e., \textless 1\%).

As no contemporaneous photometric standard was observed, the absolute flux calibration for the ERIS $L'$-band data carries an additional systematic uncertainty. We estimated the apparent $L'$-band magnitude of the central star as $\sim7.6$ based on the mean of its NEOWISE Band 1 (3.4\,$\mu$m) and Band 2 (4.6\,$\mu$m) magnitudes from \citep{Cutri2013}. While this approximation assumes minimal long-term evolution in $L'$-band flux, V960~Mon is known to be variable. We therefore assign a conservative uncertainty of ±0.2 mag to this flux calibration. Applying the measured contrast, the companion's apparent magnitude is estimated to be $L'_{\text{cand}} = 12.8 \pm 0.2$ mag.

\begin{table*}[ht]
\centering
\begin{tabular}{|c|c|c|c|c|c|c|}
\hline
\textbf{Source} & $\Delta$RA ($\arcsec$) & $\Delta$Dec ($\arcsec$) & \textbf{Separation ($\arcsec$)} & \textbf{PA ($^\circ$)} & \textbf{Contrast $\times 10^{-3}$} & \textbf{L' app. mag} \\
\hline
New Candidate & $-0.75 \pm 0.01$   & $-0.51 \pm 0.01$ & $0.89 \pm 0.01$ & $234.3 \pm 0.1$ & $8.39 \pm 0.07$ & $12.8 \pm 0.2$ \\
UCAC4 430-024261 (N) & $358.64 \pm 0.01$ & $5.31 \pm 0.01$ & $5.50 \pm 0.01$ & $340.2 \pm 0.1$ & $58.1 \pm 0.5$ & $10.7 \pm 0.1$ \\
V960 Mon NE & $355.74 \pm 0.01$ & $3.11 \pm 0.01$ & $5.28 \pm 0.07$ & $306.0 \pm 0.1$ & $1.63 \pm 0.01$ & $14.5 \pm 0.1$ \\
Source ``1" & $355.78 \pm 0.01$ & $-4.05 \pm 0.01$ & $5.87 \pm 0.05$ & $234.4 \pm 0.1$ & $4.49 \pm 0.03$ & $13.4 \pm 0.1$ \\
Source ``2" & $354.76 \pm 0.01$ & $-0.66 \pm 0.01$ & $5.58 \pm 0.01$ & $268.6 \pm 0.1$ & $0.037 \pm 0.005$ & $18.7 \pm 0.1$ \\
Source ``3" & $354.90 \pm 0.10$ & $2.37 \pm 0.10$ & $5.96 \pm 0.01$ & $296.3 \pm 0.1$ & $0.025 \pm 0.005$ & $19.1 \pm 0.1$ \\
\hline
\end{tabular}
\caption{Astrometric and photometric measurements of the detected point sources in the ERIS $L'$-band image relative to the central star.}
\label{tab:positions}
\end{table*}

\subsection{Comparison with ALMA Emission}\label{alma_comp}
The right panel of Fig.~\ref{fig:ERIS} shows a zoom-in onto V960~Mon, overlayed with ALMA band~6 continuum contours at 3, 4, and 5 $\sigma_{\rm rms}$. For the ALMA contours, we used the data set reduced with natural weights and presented in \citet{Weber2023}. We found that the nearest ALMA continuum clump lies at an offset of 187 milliarcseconds from the ERIS candidate, with coordinates $\Delta$RA = $-0\farcs562$ and $\Delta$Dec = $-0\farcs471$. This spatial offset, while small, indicates that the ERIS candidate and the ALMA clump are not spatially coincident, and cannot be due to the ALMA pointing accuracy.  

\subsection{Mass Estimation Using Evolutionary Models}

\subsubsection{Mass Estimate from ERIS in L'}


To compute the mass, we employed the \texttt{species} package \citep{2020A&A...635A.182S}, which provides a robust interface to several theoretical isochrones and evolutionary models. In particular, we used the \texttt{AMES-Cond} isochrones \citep{2001ApJ...556..357A} appropriate for low-temperature substellar objects. These models relate absolute magnitudes in specific filters (such as $L'$) to the mass and luminosity of substellar companions for a given age.

We adopted an age of $\sim$1\ Myr for the system to evaluate the properties of the detected companion using the \texttt{AMES-Cond} evolutionary models, which do not provide solutions for ages younger than 1\,Myr. This value should therefore be understood as a conservative upper limit for both the central star and the companion. Also, V960~Mon being an FU Orionis-type object is embedded in an active star-forming region, showing significant variability and signs of ongoing accretion and envelope material \citep{Kospal2015,Caratti2015}. \citet{Kospal2015} estimates the age of the central star to be $\sim$0.6\,Myr based on its spectral energy distribution, placing it among the youngest Class~II objects. 

Furthermore, the companion candidate lies along a spiral overdensity in the circumstellar environment, suggesting it may have formed through gravitational instability. Gravitational instability may also be locally triggered by late-stage infall events, particularly in dynamically active systems \citep{Kuffmeier2018}. In the case of V960~Mon, there is independent evidence for such infall from large-scale structures in its envelope and surrounding material \citep{Weber2025}. Furthermore, the fact that the candidate remains spatially coincident with a prominent spiral overdensity suggests that significant Keplerian shear or orbital motion has not yet acted to distort the structure. This implies that the formation of the companion via fragmentation is likely a recent event, consistent with a very young evolutionary stage. In this context, our use of an age of 1\,Myr taken as an upper limit for the system likely overestimates the true age of the companion and provides a conservative estimate for its mass in the evolutionary models.
 Based on the assumed age and the measured $L'$-band luminosity, and adopting a distance of $2189 \pm 281$\,pc \citep[][see also \citealt{Weber2023}]{GaiaDR3}, we estimate a mass of $\sim$660\,$M_\mathrm{Jup}$ for the companion candidate.

However, at such young ages, the luminosity of a forming object can be strongly affected by the presence of material surrounding the candidate in addition to its intrinsic mass since compact disks, such as CPDs in the case of planetary companions, are known to contribute significantly to the infrared flux, especially at $L'$-band wavelengths \citep{2015ApJ...799...16Z,2018ApJ...866...84A,2019MNRAS.487.1248S}. This means that similar high $L'$-band fluxes in young systems may arise not from the photosphere of the companion alone, but from surrounding warm dust and accreting material around a circum(sub)stellar object. In addition, if active accretion is ongoing, strong emission from the Brackett-$\alpha$ line at $4.05\,\mu\mathrm{m}$, which lies close to the $L'$-band filter, could further boost the observed flux. In such cases, the detected signal may not be purely continuum emission, but the total flux could be significantly contributed to by line emission at near and mid-infrared wavelengths arising from accretion shocks. Therefore, the derived mass of $660~M_\mathrm{Jup}$ should be interpreted with caution, as it likely reflects a total luminosity that includes circum(sub)stellar and accretion-driven components, and does not represent the true mass of a bare stellar or substellar object.

\subsubsection{Upper Mass Limit from SPHERE Non-detection}


We calculated the contrast curve from archival SPHERE data in $H$ band of the object, previously presented in \citet{Weber2023}. 
The IRDAP-pipeline \citep{vanHolstein2020} that was used to align and reduce the SPHERE data automatically calculates a contrast curve for polarized point sources.
Direct emission from the companion is expected to be unpolarized. 
Therefore, we adapted the IRDAP-pipeline to create a contrast curve for the total intensity, following the same steps as described in Appendix~E of \citet{vanHolstein2021}, i.e., calculating the standard deviation of the total intensity, $\sigma$, over several annuli centered on V960~Mon.
The contrast curve for a $5\sigma$ detection is shown in Fig.~\ref{fig:contrast}.
The spike at $\sim 0\farcs24$ is due to the contribution from the known bright stellar companion at this separation \citep{Caratti2015}.
The slight increase of the $5\sigma$ curve around $0\farcs9$ is caused by the presence of the residuals of the adaptive-optics system at this separation.

By applying the same evolutionary model approach with the \texttt{species} package (again using the \texttt{AMES-Cond} isochrones at 1~Myr), we translated the detection limit into an upper mass limit. The corresponding upper mass limit for the putative companion is $38~M_\mathrm{Jup}$. This value places an important constraint on the object's nature.

The discrepancy between the SPHERE non-detection and the ERIS detection suggests that the $L'$-band emission is likely tracing extended thermal emission from a CPD or surrounding material, rather than purely photospheric emission from a substellar object.

\section{Discussion}\label{sec:discussion}
It is worth noting that this is not the first time substellar companion candidates have been identified in young disks. Previous discoveries, such as a substellar companion potentially opening a gap in the disk around [BHB2007]-1 \citep{2021ApJ...912...64Z} and a companion candidate in the HD~169142 transition disk \citep{2014ApJ...792L..23R} highlight the aptitude of high-contrast imaging to probe the early stages of companion formation. 

The HD~169142 system shows strong evidence of planet formation, with a compact source detected at ~37 au in the gap between two bright rings, moving at the expected Keplerian velocity. Spiral features further support the presence of a planet, potentially surrounded by a circumplanetary disk \citep{Hammond2023}. A separate study identified blobs following Keplerian motion, suggesting a planet influencing the disk structure, with a mass between 1 and 4 Jupiter masses \citep{Gratton2019}.

Similarly, the PDS~70 system has provided one of the most robust cases to date, where two protoplanets, PDS~70b and PDS~70c, were detected in the gap of a protoplanetary disk \citep{2018A&A...617A..44K, 2019NatAs...3..749H}. Notably, ALMA observations have revealed a resolved CPD around PDS~70c \citep{2021ApJ...916L...2B}, marking the first confirmed detection of a CPD around a substellar companion.

 However, the scenario we present here differs in an important way: the candidate around V960~Mon is located near a structure resembling a gravitationally unstable spiral arm \citep{Weber2023}, posing the possibility that we are witnessing a substellar object forming via GI, a pathway for which direct observational evidence remains scarce.
\subsection{Interpreting the Mass Estimate from \texorpdfstring{$L'$}{L'}-band and SPHERE Non-detection}

The mass estimate of $\sim660~M_\mathrm{Jup}$ derived from the ERIS $L'$-band photometry must be interpreted with caution. While this value is formally computed using the \texttt{species} package with the AMES-Cond evolutionary models assuming a system age of 1~Myr, it reflects the total inferred luminosity at 3.8~µm under the assumption of photospheric emission alone. However, several physical considerations suggest that this estimate is likely inflated by contributions from circumplanetary material.

First, the system’s large distance (2189~pc) implies that only intrinsically bright or thermally extended sources are detectable at the measured contrast. Young, forming substellar objects are frequently surrounded by circum(sub)stellar disks \citep{2015ApJ...799...16Z, 2019MNRAS.487.1248S}, which can reprocess stellar and planetary radiation into significant mid-infrared emission. This emission would be particularly prominent in $L'$-band due to thermal dust emission and ongoing accretion processes, without necessarily reflecting a massive underlying object. Such CPD-dominated $L'$ excess has been reported in  \citet{2015ApJ...799...16Z} and \citet{2019MNRAS.487.1248S}, where a similar discrepancy between near- and mid-IR brightness was observed. Furthermore, \citet{Adams2025} derive surface density and temperature distributions for circumplanetary disks during the late stages of giant planet formation, showing that such disks can remain optically thick and thermally efficient under a wide range of infall geometries and angular momentum conditions. These conditions naturally lead to enhanced thermal emission, particularly in the mid-infrared, consistent with the elevated flux we observe in the $L'$-band. 
Also, given the candidate’s likely formation via gravitational instability (GI), we note that such objects are expected to retain significantly higher initial entropies than planets formed via core accretion (CA), owing to the absence of strong radiative losses during formation \citep[e.g.,][]{Mordasini2013}. Evolutionary models typically assume “hot start” conditions, but even these may underestimate the true luminosity of a GI-formed object. Therefore, for a given observed luminosity, GI objects could have lower masses than predicted by standard hot-start models. This reinforces our conclusion that the estimated mass of the candidate represents an upper limit.

Second, the SPHERE $H$-band non-detection provides a much lower upper mass limit of $\sim38~M_\mathrm{Jup}$. This suggests that the object’s photosphere is not directly detected at shorter wavelengths, either due to intrinsic faintness or obscuration. The co-location of the candidate with extended polarized light emission (Section~\ref{subsec:pollight}) supports the idea that the companion is embedded in a dusty environment, likely contributing to extinction in the $H$-band. The same dust may enhance thermal emission in the $L'$-band, further skewing the mass estimate derived from evolutionary models that do not account for disk contributions.

Taken together, this implies that the ERIS-based mass is likely an upper limit, and that the true mass of the object may lie well below the formal estimate, potentially in the planetary or low brown dwarf regime. Future spectroscopic analysis or HCI at longer wavelengths will be key to disentangling these components and obtaining a more accurate mass constraint.

To further assess whether the non-detection in the $H$-band could be explained by dust extinction alone, we estimated the required visual extinction assuming the candidate is a young low-mass stellar object (age $\sim$1~Myr) with an intrinsic color of $(H - L')_0 \sim 0.7$~mag, consistent with empirical values for late-M type stars \citep{Luhman2010, Pecaut2013}. Given the measured $L'$-band magnitude of 12.8 and a conservative upper limit of $H > 20.12$~mag, derived from the 2MASS $H$-band magnitude of the central star ($H = 10.12$~mag; \citealt{Cutri2023}) and a 5$\sigma$ contrast limit of $\Delta H > 10$~mag (Figure~\ref{fig:contrast}) at the companion's separation, the observed color is $(H - L') > 7.32$~mag. This implies a color excess of $E(H - L') > 6.62$~mag. Using a standard extinction law \citep{Cardelli1989,Indebetouw2005}, where $A_H / A_V = 0.175$ and $A_{L'}/A_V = 0.058$, we find that reproducing the observed color would require an extinction of $A_V > 56.6$~mag. Such a high value suggests that the object is deeply embedded in circum(sub)stellar material. This supports our interpretation that the observed $L'$-band flux is likely dominated by emission from a compact dusty envelope or a circumplanetary disk rather than from a photosphere alone.

\subsection{On the Offset Between ALMA and ERIS Emission}

The spatial offset of approximately 187 milliarcseconds between the ERIS-detected $L'$-band candidate and the nearby ALMA Band~6 continuum clump raises important questions about the physical relationship between the two components. While the candidate is clearly visible in the ERIS image as a compact point source, the ALMA emission appears as an extended clump, offset to the northeast.

A similar observational framework has been studied in the case of the PDS~70 system. In particular, \citet{2021ApJ...916L...2B} reported the direct detection of a CPD around PDS~70c using ALMA at 855~µm, revealing spatially resolved continuum emission centered on the planet. In that system, the ALMA-detected dust emission was found to be co-located with the planet’s position as derived from H$\alpha$ observations with VLT/MUSE \citep{2019NatAs...3..749H}, providing strong evidence that the submillimeter continuum traces a compact CPD coincident with the forming planet. However, for PDS~70b, \citet{2019ApJ...879L..25I} reported a submillimeter continuum source offset from the planet’s infrared position, underscoring that such offsets can occur and may reflect distinct spatial components of the circumplanetary environment.

In contrast, the situation in V960~Mon may be more complex. The observed offset could arise from a combination of factors. First, the ALMA emission may be tracing an extended dust structure—possibly a circum(sub)stellar, spiral arm fragment, or envelope—that is spatially disconnected from the infrared-bright core detected in $L'$-band. Second, astrometric uncertainties, especially due to different observing epochs, may contribute to the apparent offset. The ALMA Band~6 observations were obtained in 2016 as part of program 2016.1.00209.S (PI: Takami) and reanalyzed by \citet{Weber2023}, while the ERIS $L'$-band data were acquired approximately eight years later. Given the central star's mass of \(0.6~M_\odot\) \citep{Kospal2015} and a projected separation of about 2000~au, the expected orbital motion over this time baseline is negligible. Thus, the offset is unlikely to result from physical motion and more likely reflects astrometric calibration differences or differences in the emission morphology at submillimeter and infrared wavelengths.

If the ALMA clump and the ERIS point source are physically associated, the offset may point to a scenario where the candidate is still embedded within a dust-rich environment. In this case, the $L'$-band detection could correspond to the heated inner region of a circum(sub)stellar or the planet’s photosphere, while the ALMA emission traces colder dust in the outer disk. The discrepancy in positions would then reflect the stratified structure of circum(sub)stellar material rather than a true separation between two independent sources.

Follow-up observations at higher angular resolution and across multiple wavelengths—especially in the mid-infrared and submillimeter regimes will be essential to confirm whether the ALMA clump and the ERIS companion represent different parts of the same physical system or separate components altogether.

\subsection{SPHERE polarized light signal}\label{subsec:pollight}
\begin{figure}
    \centering
    \includegraphics[width=\columnwidth]
    {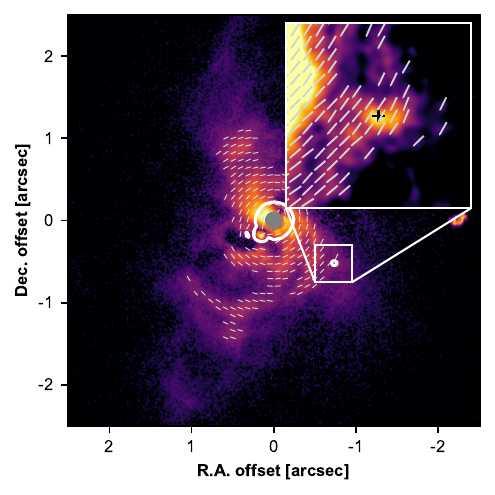}
    \caption{SPHERE/IRDAP polarised light image of V960~Mon in H-band \citep[see][]{Weber2023}.
    The white contours show the $L'$-signal.
    The grey circle delineates the coronagraph used in the observation.
    The white bars show the AoLP of the polarized light.
    The inset is at increased contrast and centered on the location of the ERIS candidate, marked by a black cross.}
    \label{fig:polarized}
\end{figure}

V960~Mon has been observed in dual-beam polarimetric imaging mode \citep[DPI,][]{deBoer2020,vanHolstein2020} for which the polarized light image was presented in \citet{Weber2023}.
Interestingly, we find a polarized light signal at the location of the ERIS candidate, as shown in Fig.~\ref{fig:polarized}.
The inset shows that there is extended flux clearly attributed locally to the candidate.
Is this flux emission from the $L'$-band object?
The initial emission of a compact object is expected to be unpolarized.
To produce a polarized light signal, the light needs to become polarized along its path.
In principle, there are several possibilities: the light can be scattered by clouds in the candidate's atmosphere \citep[e.g.][]{Stolker2017}, scattered by a highly inclined circumplanetary or circum-substellar disk \citep[e.g.][]{vanHolstein2021}, or scattered along the line of sight in the interstellar medium.
However, we point out two reasons why it is unlikely that this flux originates from the candidate.
First, the polarized signal does not resemble the PSF of the observation.
Both a disk surrounding the candidate and atmospheric clouds would be unresolved and, therefore, should resemble a point source.
For the second reason, we examine the white bars in Fig.~\ref{fig:polarized}, which show the Angle of Linear Polarization (AoLP), defined as
\begin{equation}
    \mathrm{AoLP} = \frac{1}{2} \arctan\left(\frac{U}{Q}\right)\,.
\end{equation}
The AoLP is an important measure when assessing the light source that produces the scattered light, as it is expected to be perpendicular to the direction from which the light originated.
In Fig.~\ref{fig:polarized}, we observe that the AoLPs at the candidate's location are perpendicular to the direction towards the primary, indicating that we see light originating from the primary that is scattered by small dust grains at the candidate's location.
Therefore, we suggest that the candidate is partly embedded in a cloud of gas and small dust grains, attenuating its intrinsic emission.

\subsection{\texorpdfstring{$H$}{H}-band total intensity}\label{subsec:totalIntesity}

As argued in the previous section, direct emission from the candidate is expected to be unpolarized.
Therefore, we inspected the total intensity image of the SPHERE/IRDIS $H$-band observation, which is produced by calculating the double sum of the individual DPI-frames \citep{vanHolstein2020}.
At the $L'$-candidates location, we do not find any signal above the local noise level in $H$-band total intensity.
The detection is impeded by the candidate's co-location with the primary's AO residuals at $1.625\,$µm.
\begin{figure}
    \centering
    \includegraphics[width=\columnwidth]{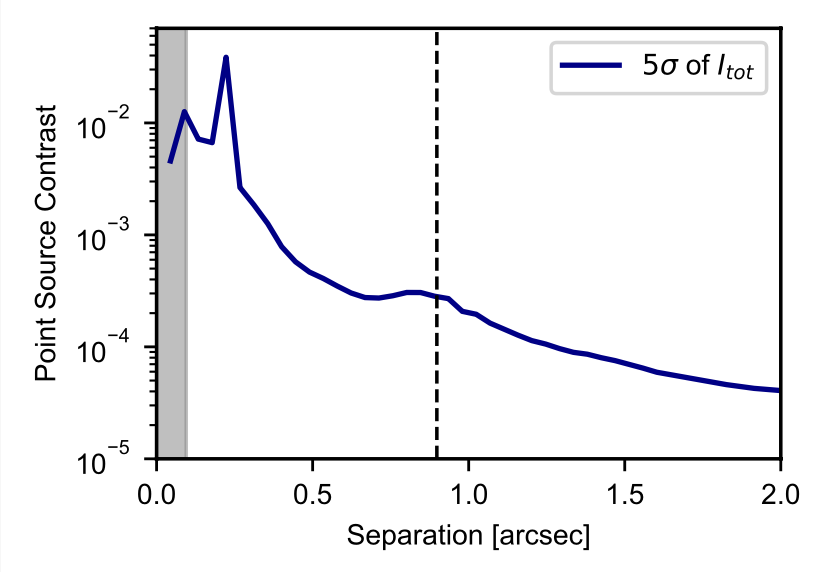}
    \caption{Contrast curve for the total intensity from SPHERE  $H$-band. The vertical dashed line shows the separation of the $L'$-band candidate. The grey-shaded area shows the separation covered by the coronagraph.}
    \label{fig:contrast}
\end{figure}
In Fig.~\ref{fig:contrast}, we show the $5\sigma$ contrast curve of the total intensity $H$-band observation.
The non-detection puts constraints on the candidate's nature, or promotes the notion of flux attenuation through extinction. 
As already indicated in Sec.~\ref{subsec:pollight}, there are signs that the candidate is deeply embedded in a cloud of small dust grains, hindering its light from escaping at small wavelengths.
If there is indeed strong extinction along the line-of-sight, the non-detection in $H$-band is inconclusive for the candidate nature, and observations at longer wavelengths are required
We also note that several sources identified in our ERIS $L'$-band image, specifically the companion candidate, the northern component (V960 Mon N, also known as UCAC4430-024261), the northeastern source (NE), Source 1 and  Source 2—are visible in the reduced SPHERE $H$-band total intensity image. While these sources were not formally identified in the original SPHERE data analysis, they are consistently detected at similar locations. For clarity, we provide an annotated version of the SPHERE image in the Appendix (\ref{fig:appendix_hband}).

\subsection{Probability of a Background Contaminant}

To assess the likelihood that the detected candidate around V960 Mon is a chance-aligned background star rather than a bound companion, we performed a TRILEGAL \citep{girardi2005} simulation centered on the coordinates of V960 Mon (RA = 06:47:02.3, Dec = $-06^\circ 42' 43.8''$, $l = 217.4942^\circ$, $b = -0.0691^\circ$). We simulated a field of area 0.001 deg$^2$, adopting the ``2MASS + Spitzer” filter set and using IRAC 3.6 $\mu$m as a proxy for our $L'$-band.  

We set the faint limit to the candidate’s apparent magnitude, $L'_{\rm cand} = 12.8$ mag (derived from the host star’s $L'_{*}=7.6$ mag and the measured contrast).  The TRILEGAL output returned $N = 56$ stars with $[3.6]\le12.8$ mag in the 0.001 deg$^2$ field, corresponding to a surface density  
\[
  \Sigma \;=\;\frac{N}{0.01~\mathrm{deg}^2}
  \;=\;5{,}600\ \mathrm{stars\,deg}^{-2}.
\]
The candidate lies at a projected separation $r = 0.898''$, defining a circular search area  
\[
  A \;=\;\pi\,r^2
    \;=\;\pi\,(0.898'')^2
    \;\approx\;2.53\ \mathrm{arcsec}^2
    \;=\;7.06\times10^{-7}\ \mathrm{deg}^2.
\]
Thus the probability of a background star of similar brightness falling by chance within this region is  
\[
  P_{\rm bg} \;=\;\Sigma \,A
  \;=\;5{,}600\times7.06\times10^{-7}
  \;\approx\;3.95\times10^{-3}
  \;\simeq\;0.4\%.
\]
Although low, this non-zero probability,combined with the proximity of the candidate to ALMA continuum clumps (offset by only 187 mas; Section~\ref{alma_comp}), motivates a second‐epoch observation with ERIS to test for common proper motion and confirm bound status.

\FloatBarrier
\section{Conclusions}\label{sec:conclusions}

We presented high-contrast VLT/ERIS $L'$-band imaging of the young eruptive star V960 Mon, unveiling the detection of a new substellar companion candidate at a projected separation of $0.898 \pm 0.01 \arcsec$ and a contrast of $(8.39 \pm 0.07) \times 10^{-3}$. The candidate is found close to structures previously identified in ALMA continuum emission and lies within a region exhibiting extended polarized light, suggesting it is deeply embedded in dust.

Using evolutionary models at an assumed age of 1~Myr, we estimate a mass of $\sim660$~$M_\mathrm{Jup}$ based on the $L'$-band brightness. However, the non-detection in archival SPHERE $H$-band imaging places a much stricter upper mass limit of $\sim38$~$M_\mathrm{Jup}$, indicating that the observed infrared flux is likely dominated by circumplanetary or circum(sub)stellar dust emission rather than photospheric light alone. These results highlight the critical role of circumplanetary material in interpreting infrared detections in young, embedded systems.

The spatial offset between the ERIS detection and the nearby ALMA clump, although small, suggests a complex circumstellar environment. The candidate's co-location with polarized light emission further supports the notion of an embedded object obscured by dust, possibly in an early formation stage.

If confirmed, this source may represent the first direct detections of a substellar object forming via gravitational instability in a massive, young disk---providing valuable observational support to theoretical models of GI-driven planet and brown dwarf formation. Future multi-wavelength observations, including mid-infrared and submillimeter imaging, will be essential to confirm the nature of the candidate, constrain its mass more accurately, and disentangle the contributions of circumplanetary emission and extinction.

\section*{Acknowledgments}

We thank the anonymous referee for the construcive comments,
which have helped improve the Letter. 
This paper is based on observations collected at the European Organisation for Astronomical Research in the Southern Hemisphere under ESO programmes 098.C-0422(B) and 112.25PY.001
This paper also makes use of the following ALMA data: ADS/JAO.ALMA\#2016.1.00209.S. ALMA is a partnership of ESO (representing its member states), NSF (USA), and NINS (Japan), together with NRC (Canada), NSTC and ASIAA (Taiwan), and KASI (Republic of Korea), in cooperation with the Republic of Chile. The Joint ALMA Observatory is operated by ESO,

The authors acknowledge the support from the Millennium
Nucleus on Young Exoplanets and their Moons (YEMS),
ANID — Center Code NCN2021\_080 and NCN2024\_001. A.Z. acknowledges Fondecyt Regular grant 1250249. P.W. acknowledges support from FONDECYT grant 3220399. L.A.C. acknowledges support from ANID, FONDECYT Regular grant No. 1241056, Chile. S.P. acknowledges support from Fondecyt Regular grant 1231663.  V.R. acknowledges the support of the European Union’s Horizon 2020 research and innovation program and the European Research Council via the ERC Synergy Grant “ECOGAL” (project ID 855130).

\appendix
\section{SPHERE H-band Total Intensity Image}

\begin{figure*}[ht]
    \centering
    \includegraphics[width=0.7\columnwidth]{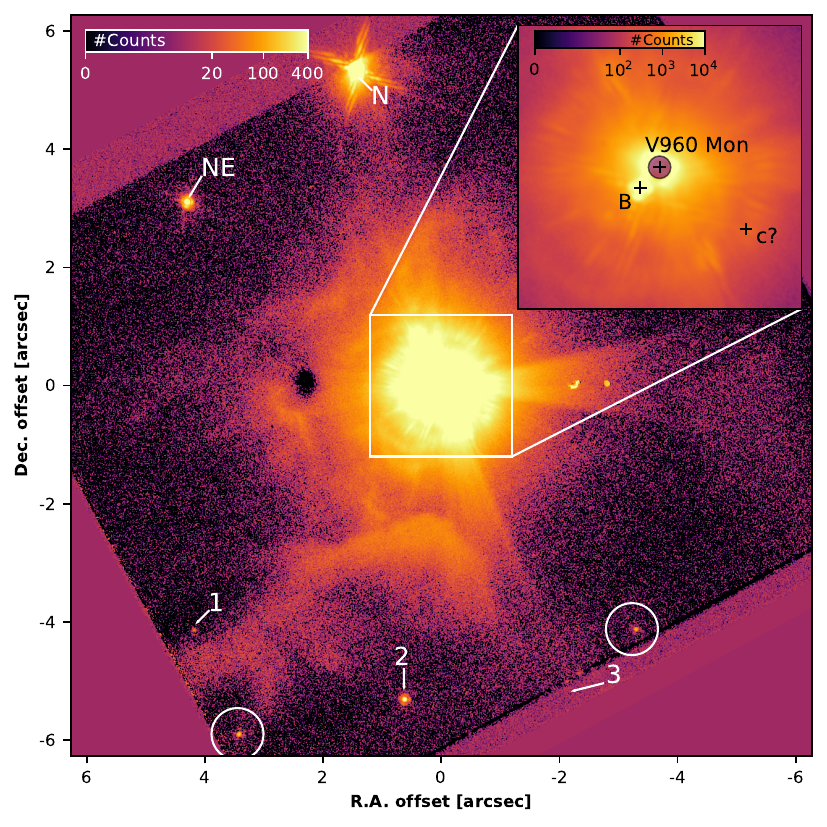}
    \caption{ SPHERE/IRDIS $H$-band ($\lambda_{\rm obs}=1.625\,\mu{\rm m}$) total intensity image of V960~Mon \citep[cf.][]{Weber2025}.
    The main panel saturates the inner region, to highlight the detection of Source~1 and Source~2, as well as the northern and north eastern components in $H$-band.
    Source~3 falls just outside the SPHERE detector.
    In $H$-band there are two more point sources visible in this region, which we mark with white circles.
    Note, that the two additional dots east of V960~Mon are detector artifacts. 
    The inset focuses on the inner region on a different scale.
    Here, sources are marked by black crosses.
    V960~Mon itself is behind the coronagraph, marked by a purple circle.
    We want to highlight the detection of the ‘B' component and especially the non-detection of the ERIS candidate (c) that is at the base of our study.}

    \label{fig:appendix_hband}
\end{figure*}

Figure~\ref{fig:appendix_hband} shows the SPHERE/IRDIS $H$-band total intensity image of V960~Mon.
This image is based on the same dataset presented in \citet{Weber2025} and was produced using the IRDAP-pipeline \citep{vanHolstein2020}.
All the sources detected by ERIS also appear in the SPHERE image, except for Source~3, which is outside the SPHERE detector, and more importantly, the candidate southeast of the primary.

\FloatBarrier
\bibliography{references}{}
\bibliographystyle{aasjournal}

\end{document}